\documentstyle[preprint,aps]{revtex}
\newcommand{\be}{\begin{equation}}
\newcommand{\ee}{\end{equation}}
\newcommand{\bea}{\begin{eqnarray}}
\newcommand{\eea}{\end{eqnarray}}
\newcommand{\ba}{\begin{array}}
\newcommand{\ea}{\end{array}}     
\newcommand{\nn}{\nonumber \\}
\newcommand{\half}{\frac{1}{2}}
\newcommand{\bb}{\bibitem}

\begin{document}
\draft 
\preprint{}

\title{CANONICAL AND FUNCTIONAL SCHR\"ODINGER QUANTIZATION OF
  TWO--DIMENSIONAL DILATON GRAVITY} 

\author{S. Cassemiro F. F.$^1$ \\ and \\ Victor
  O. Rivelles$^2$}  \address{ Instituto de F\'\i sica, 
  Universidade de S\~ao Paulo \\  Caixa Postal 66318, 05315--970, S\~ao
  Paulo, SP, Brazil \\ $^1$ E-mail: figueire@fma.if.usp.br \\ $^2$ E-mail:
  rivelles@fma.if.usp.br }

\maketitle

\begin{abstract}
We discuss the relation between canonical and Schr\"odinger
quantization of the CGHS model. We also discuss the situation when
background charges are added to cancel the Virasoro anomaly. 
New physical states are found when the square of the background charges
vanishes.   
\end{abstract}

\narrowtext
\newpage

The quantization of reparametrization invariant theories is an open
problem with many unanswered questions. Even in two space--time
dimensions many delicate questions remain without a clear answer. In
the last few years much effort has been  spent in the study of several
two--dimensional dilaton gravity models as prototypes of
reparametrization invariant theories. In particular the string
inspired CGHS (Callan, Giddings, Harvey and Strominger) model
\cite{CGHS}  was intensively investigated. 

The CHGS model consists of a particular coupling of 
two--dimensional gravity and a dilaton field. It is described by the
action
\be \label{CGHS}
S = \int d^2 x \: \sqrt{-g} \: e^{-2\phi} \left( R + 4 g^{\mu\nu}
  \partial_\mu \phi \partial_\nu \phi - \Lambda \right), 
\ee
where $R$ is the scalar curvature, $\phi$ is the dilaton and $\Lambda$
the cosmological constant. Matter fields can also
be added but we will not consider them. The interest in the CGHS model
stem from the fact that the model allows black hole formation and
Hawking radiation at the semi--classical level \cite{semiclassical}. 
From a more formal point of
view the CGHS model has also interesting properties. It 
can be reformulated as a topological field theory of the BF type with
the gauge group being  the extended Poincar\'e group
\cite{Cangemi-Jackiw}. Its supersymmetric version is also known
\cite{susy}. It is possible to go beyond the semi--classical
level and to quantize the model without any approximation. Most
surprisingly, the reparametrization constraints of the CGHS model can
be set in a quadratic form after a conformal transformation followed
by a canonical transformation \cite{Mikovic,canonical}.  
In this form there are two scalar fields $r^a(x), a=1,2$ corresponding
to combinations of the gravity and dilaton fields with a vanishing
Hamiltonian which is characteristic of reparametrization invariant
theories. The first order form of the Lagrangian in the absence of
matter and with the cosmological constant $\Lambda = 1$ is 

\begin{equation} \label{1}
L = \pi_a {\dot r}^a + \lambda_0 H_0 + \lambda_1 H_1,
\end{equation}
where $\pi_a(x)$ is the canonical momentum of $r^a(x)$ and
$\lambda_a(x)$ are 
the Lagrange multipliers which implement  the
reparametrization constraints 

\begin{eqnarray} \label{2}
{H}_0 &=& {1\over{2}} ( \pi^a \pi_a + r^{\prime a} r^\prime_a ), \nn
{H}_1 &=& \pi_a r^{\prime a}. 
\end{eqnarray}
In our notation a dot (dash) indicates differentiation with respect to time
(space). The ``internal'' indices $a, b, \dots$ are raised and lowered with a
Minkowskian metric $\eta_{ab} = diag(1,-1)$ (not the space--time
metric) so that the canonical variables $\pi_a, r^a$ appear in an
indefinite quadratic form in Eqs.(\ref{1}, \ref{2}). We will then say that
the field $r^0(x)$ has positive signature while $r^1(x)$ has negative
signature. Note that the
constraints Eqs.(\ref{2}) are just the components of the
energy--momentum tensor of two massless scalar fields with opposite
signature. 

The theory described by Eqs.(\ref{1}, \ref{2}) looks very simple since
there are no interaction terms. In the gauge $\lambda_0=0,
\lambda_1=1$ it describes two massless scalar fields with opposite
signature and with a vanishing energy--momentum tensor. We would
expect that the physical states should be the direct product of states
for each degree of freedom separately. However, there are subtle
correlations due to the constraints Eqs.(\ref{2}) and the Hilbert
space has not a direct product structure. 

The canonical quantization of the theory is upset with anomalies. Due
to the normal ordering in the constraints Eqs.(\ref{2}) there appears 
the well known Virasoro anomaly in the algebra of the
energy--momentum tensor. It is 
possible to cancel the anomaly in three different ways and the resulting
quantum theories are not equivalent. The first 
possibility is to make a 
non conventional choice of the vaccum for one of the fields
$r^a(x)$ \cite{Cangemi-Jackiw-Zwiebach}. It is non conventional in the
sense that the usual creation and 
annihilation operators have their role reversed. For this field the resulting
central charge changes sign. Then the overall central charge vanishes
and no anomaly is present. In the second possibility we add background
charges to the scalar fields \cite{Cangemi-Jackiw-Zwiebach}. By choosing
appropriately the value of the background charges the anomaly can be
made to cancel. Ghosts can also be added in this case. 
The third procedure consists in 
modifying the constraints in order to cancel the anomaly
\cite{Kuchar} but we will not take this route here. 

In this paper we will concentrate on the first and second
procedures. We will find new physical states in the presence of
background charges. 

The usual way to incorporate background charges is to consider an
improved energy--momentum tensor. To derive this improved
energy--momentum tensor consider the Lagrangian of a free massless scalar
field $\phi$ with a surface term linear in the field $Q \Box \phi$,
where $Q$ is the background charge. From this Lagrangian  
we can find the appropriate energy--momentum tensor. When this is done
for the fields $r^a$, taking into account their signature, we find 

\begin{equation} \label{4}
T_{\mu\nu} = \half \partial_\mu r^a \partial_\nu r_a - \frac{1}{4}
\eta_{\mu\nu} \partial^\rho r^a \partial_\rho r_a + \half Q_a
\partial_\mu \partial_\nu r^a - \frac{1}{4} \eta_{\mu\nu} Q_a \Box
r^a.
\ee
The constraints ${H}_0 = ( T_{++} + T_{--})/2 $ and ${H}_1 = ( T_{++}
- T_{--} )/2 $ are now 

\be \label{5}
H_0 = {1\over{2}} ( \pi^a \pi_a + r^{\prime a} r^\prime_a ) + Q_a
r^{\prime\prime a}, 
\ee
\be \label{5a}
H_1 =  \pi_a r^{\prime a} + Q_a \pi^{\prime a}.
\ee
As usual the Poisson bracket algebra of the constraints acquires a
classical central charge due to the surface term

\bea \label{6} 
\{ H_0 (x), H_0 (y) \} &=& \left( H_1 (x) + H_1 (y) \right)
\delta^\prime ( x-y ), \nn
\{ H_0 (x), H_1 (y) \} &=& \left( H_0 (x) + H_0 (y) \right)
\delta^\prime ( x-y ) - Q^a Q_a \delta^{\prime\prime\prime} ( x-y ),
\nn
\{ H_1 (x), H_1 (y) \} &=& \left( H_1 (x) + H_1 (y) \right)
\delta^\prime ( x-y ).
\eea
Therefore the new constraints have a first class algebra only if $Q_a
Q^a = 0$. A careful analysis of the canonical transformation which brings
the original CGHS model Eq.(\ref{CGHS}) (written in terms of the
dilaton and gravity 
fields) to the quadratic form Eq.(\ref{1}) (written in terms of $r^a$)
shows that the two Lagrangians differ by a surface term
\cite{Mikovic}. This surface term can be written in the form $Q_a \Box
r^a$ with $Q_a Q^a=0$. This is necessary to
retain the reparametrization symmetry of the original model. However,
since the quantum theory is afflicted with anomalies we are allowed to
modify it by adding a Wess--Zumino field to cancel the
anomaly. Since in two--dimensions a scalar field can serve as its own
Wess--Zumino field we can add an improvement term to the constraints
with an appropriate value of $Q_a Q^a$ to cancel the anomaly. So, in
general, 
$Q_a Q^a$ will no longer vanish and the classical theory will loose
reparametrization invariance. However, it will be recovered at the quantum
level. 

Before performing the canonical quantization with the new constraints 
let us first consider a single massless scalar field. In the canonical
approach it has an expansion in terms of 
Fock space operators $a^\dagger(k)$ and $a(k)$  associated with
particles of positive and negative energy respectively. The
conventional vaccum is defined as $a(k) |0> = 0$ so that the Hilbert
space is positive definite and the 
energy of the states is also positive. This gives rise to a
central charge $c=1$ in the energy--momentum tensor algebra when
normal ordering is taken into account. An alternative choice for the
vaccum is to take  $a^\dagger(k) |0> = 0$. In this case the Hilbert
space is no longer positive definite, the  
energy of the states is negative and the central charge is
$c=-1$. For conventional theories this choice of the vaccum is not
allowed. 

Let us now consider a single scalar field with negative signature. In
the canonical 
approach there is a crucial change of sign in the canonical momentum
which leads to a change of sign in the algebra of creation and
annihilation operators. Now if the vaccum is defined as $a(k) |0> = 0$
then the 
Hilbert space is not positive definite but the energy of the states is
positive and the central charge is $c=1$. For the other choice of the
vaccum $a^\dagger(k) |0> = 0$ the Hilbert space is positive definite,
the energy is negative and $c=-1$. Then the quantum theory of a scalar
field with negative signature has troubles for any choice of the
vaccum.  

When a background charge $Q$ is added its effect is just to shift the
value of the central charge. A short calculation shows that for the
conventional scalar field we 
have for the usual choice of the vaccum $a(k) |0> = 0$ the value $c =
1 + 12 \pi Q^2$ while for the vaccum $a^\dagger(k) |0> = 0$, $c=-1 + 12
\pi Q^2$. For the scalar field with negative signature and vaccum
$a(k) |0> = 0$ we have  $c = 1 -12 \pi Q^2$, while for the vaccum
$a^\dagger(k) |0> = 0$ we find $c=-1 - 12 \pi Q^2$. This is summarized
in Table \ref{fig1}. 

As remarked before the CGHS model written in the form Eq.(\ref{1})
involves two free massless scalar fields with opposite signature as
can be seen when the gauge ${\lambda_0} = 1, {\lambda_1} = 0$ is
choosen. Then canonical quantization allows several possibilities for
the vanishing of the total central charge. If no background charges
are present we can achieve $c=0$ by choosing the vaccum $a_0(k) |0> =
a^\dagger_1 (k)|0> = 0$. Note that since our Hamiltonian is zero we
have no troubles with the positivity of the energy. If background
charges with $Q_a Q^a = 0$ are present we must do the same vaccum  
choice. If the background charges have $Q_a Q^a \not= 0$ then
the vanishing of the central charge requires $Q_a Q^a = \pm 1/(6 \pi)$
depending on which vaccum is choosen. There are two possiblities:
$a_0(k) |0> = a_1 (k) |0> = 0$ or $a^\dagger_0 (k) |0> = a^\dagger_1
(k) |0> = 0$. Either possibility is troublesome since positivity of
the Hilbert space is compromised. We will also meet difficulties for
the case $Q_a Q^a \not= 0$ in the Schr\"odinger representation. These
results are presented in Table \ref{table2}

Physical states have been explicitely constructed for the case $Q_a
=0$ \cite{Cangemi-Jackiw-Zwiebach}. For the case $Q_a \not= 0$ they
have been found when the topology of space--time is non--trivial. We
will comment on this at the end of the paper.  Ghosts can also be
added. They simply change the value of the background charges and the
same analysis carries through.  

We now consider the Schr\"odinger representation.  The
Scr\"odinger functional $\Psi$ is a functional of $r^a$,  $\Psi(r^a)$,
and $\pi_a$ is realized as a functional derivative $\pi_a(x) = - i \delta
/ \delta r^a(x)$.  In the 
Schr\"odinger representation there is no normal products to be taken
into account. The only source of ambiguity is in the operator ordering. So
the questions about the value of the central charge are difficult to
be posed in this formalism. The relevant point here is whether there
is a first class algebra of quantum constraints so that physical
states can be properly defined. 

When the Poisson bracket algebra of the constraints Eqs.(\ref{6}) is
replaced 
by the respective commutator algebra we obtain the same central charge
proportional to $Q_a Q^a$. The algebra of the constraints is not
first class and physical states can not be defined unless
$Q_a Q^a=0$. Alternatively we could try to modify the constraints to take
normal ordering into account in order to recover a first class 
algebra. So let us consider the effect of normal ordering in each term
of the constraints. Let us assume again that we have a single scalar
field $\phi$. Assuming that $\phi(x)$ and $\pi(x)$ have canonical
commutation relations we find that

\be \label{8}
: \phi^\prime(x) \pi(y) : \: = \phi^\prime(x) \pi(y) - \frac{i}{2}
\delta^\prime(x-y),
\ee
for any choice of the vaccum and for any signature of the field. This
means that

\be \label{9}
: H_1(x) : \: = r^{\prime a} \pi_a + Q_a \pi^{\prime a} -
i \lim_{y \rightarrow x} \delta^\prime(x-y), 
\ee
which does not depend on which vaccum is choosen. This is the
same ambiguity that we find if we consider the operator ordering in
$H_1$. Since $\pi_a$ and $r^a$ have canonical commutation relations
there is an ambiguity in the term $ \pi_a r^{\prime a}$ in Eq.(\ref{5a})
with the same form as in Eq.(\ref{9}). Then the coefficient of the
$\delta^\prime(x-y)$ term is not fixed. For each choice of this
coefficient we have an operator ordering prescription. This is also
consistent with the commutator algebra of the constraints. It is
independent of the value for this coefficient as it is easily
verified. 

Let us now consider the effect of normal ordering in $H_0$. If the 
field $\phi$ has positive signature 

\be \label{10}
: \pi(x) \pi(y) : \: = \pi(x) \pi(y) \mp \half \omega(x-y),
\ee
where

\be \label{10a}
\omega(x-y) = \frac{1}{2\pi} \int dk \: |k| e^{i k ( x-y )}.
\ee
The upper sign in Eq.(\ref{10}) is for the usual vaccum $a|0>=0$
while the lower sign is for the unusual vaccum $a^\dagger|0>=0$. If
the field $\phi$ has negative signature then 

\be \label{11}
: \pi(x) \pi(y) : \: = \pi(x) \pi(y) \pm \half \omega(x-y),
\ee
with the upper (lower) sign for the usual (unusual) vaccum. The same
structure holds for $\phi^\prime(x) \phi^\prime(y)$. Therefore we find
that  

\be \label{12}
: H_0 : \: = {1\over{2}} ( \pi^a \pi_a + r^{\prime a} r^\prime_a ) + Q_a
r^{\prime\prime a} + \frac{c}{2} \lim_{y \rightarrow x} \omega (x-y),
\ee
where $c=0, \pm 2$ is the sum of the central charges of $r^0$ and
$r^1$. This takes into account all possible choices of the
vaccum. 

Therefore the constraints in the form Eqs.(\ref{9},\ref{12})
are now first class and must be used for seeking solutions in the
Schr\"odinger representation. The equations for the physical states
are then  

\be \label{13}
r^a(x) \frac{\delta \Psi}{\delta r^a(x)} + Q^a \left( \frac{\delta
   \Psi}{\delta r^a(x)} \right)^\prime - \alpha \lim_{y \rightarrow
  x} \delta^\prime (x-y) \Psi = 0,
\ee

\be \label{14}
\half \left( - \frac{\delta^2 \Psi}{\delta r^a(x) \delta r_a(x)} +
  r^{\prime a}(x) r^\prime_a (x) \Psi \right) + Q_a r^{\prime \prime
  a}(x) \Psi  + \frac{c}{2} \lim_{y \rightarrow x} \omega (x - y) \Psi
= 0, 
\ee
where $\alpha$ is a constant which will select the operator ordering
prescription. The simplest choice is to take $\alpha=0$ and, as we will
see, we can find solutions with this prescription. So we will adopt it
from now on. 

We will now look for the vaccum state in the Schr\"odinger
representation. This is most easily done going to the canonical
formalism and expressing the creation and annihilation operators in
terms of $r^a$ and $\pi_a$. We find 

\be \label{15}
a_a(k) = \frac{1}{\sqrt{4\pi |k|}} \int dx \: \left( |k| r^a(x) \pm i
  \pi_a(x) \right),
\ee
where the upper sign holds for $a=0$ and the lower sign for $a=1$. As
we have seen before the vaccum is the same for $Q_a = 0$ and $Q_a Q^a = 0$
cases. It is defined by

\be \label{16}
a_0 (k) \Psi_{vaccum} = a^\dagger_1 (k) \Psi_{vaccum} = 0, 
\ee
and the solution is known \cite{Cangemi-Jackiw-Zwiebach}

\be \label{17}
\Psi_{vaccum} =  {\det}^\half \left( \frac{\omega}{\pi} \right)  \exp
\left[ -\half \int dx \: dy \: \left(  r^0(x) \omega(x-y) r^0(y) + r^1(x)
    \omega(x-y) r^1(y) \right) \right].
\ee
Since $\omega(x-y)$ Eq.(\ref{10a}) has a positive kernel this vaccum is
normalizable. For the case $Q_a Q^a \not= 0$ the vaccum would be defined by 

\be \label{18}
a_0 (k) \Psi_{vaccum} = a_1 (k) \Psi_{vaccum} = 0
\ee
or 

\be \label{19}
a^\dagger_0 (k) \Psi_{vaccum} = a^\dagger_1 (k) \Psi_{vaccum} = 0.
\ee
These equations do not have normalizable solutions. The solution of
Eq.(\ref{18}), for example, is given by \cite{Cangemi-Jackiw-Zwiebach}

\be \label{20}
\Psi = \exp \left[ -\half \int dx \: dy \: \left(  r^0(x) \omega(x-y) r^0(y) -
    r^1(x) \omega(x-y) r^1(y) \right) \right].
\ee
The minus sign in front of the second term makes the wave functional
non normalizable since the kernel is positive. This shows that there
is no vaccum state for $Q_a Q^a \not= 0$ in the Schr\"odinger
representation. In the canonical analisys we found that in this case
the Hilbert space is not positive definite. 

We now look for physical states in Eqs.(\ref{13},\ref{14}). For the
case $Q_a=0$ they are already known 
\cite{Cangemi-Jackiw-Zwiebach}. Taking $Q_a=c=0$ in
Eq.(\ref{13},\ref{14}) we obtain 

\be \label{21}
\Psi_{Q=0}  = \exp \left( \pm \frac{i}{2} \int dx \: \epsilon_{ab}
  r^a(x) r^{\prime b}(x) \right).
\ee
For the case $Q_a Q^a=0$ we still have $c=0$ in Eq.(\ref{14}) and we find

\be \label{22}
\Psi_{Q_a Q^a=0}  = \exp \left[ \pm \frac{i}{2} \int dx \: \epsilon_{ab}
  r^a(x) r^{\prime b}(x) \pm i \int dx \: \epsilon_{ab} Q^a r^{\prime
    b}(x) \ln \left( \epsilon_{cd} Q^c r^{\prime d}(x) \right) \right].
\ee
This solution reduces to the former solution when $Q_a=0$. It is
possible to rewrite it in many other forms thanks to the identity 

\be \label{23}
\epsilon_{ab} Q^a r^{\prime b} \: Q_c r^{c} = \epsilon_{ab}
Q^a r^{b} \: Q_c r^{\prime c}, 
\ee
which holds for $Q_a Q^a=0$. 
Another form for the argument of the second term in the exponential in
Eq.(\ref{22})  could be $\epsilon_{ab} Q^a r^{\prime b} \ln \left( Q_c
  r^{\prime c} \right)$. It is easier to work with the form given in
Eq.(\ref{22}). We suspect that the Fock space state corresponding to
the wave functional Eq.(\ref{22}) will have a very cumbersome form due
to the presence of the logarithmic term. 

These new physical states Eq.(\ref{22}) have not been previously
identified in the BRST formulation \cite{circle}. The reason is that
the general BRST techniques \cite{BRST} that have been applied to the
problem hold only when the background charges have $Q_a Q^a
\not=0$ which is not our case.

As we have seen there are no physical states for $Q_a Q^a \not=
0$. However, they have been found when the space--time topology is $R
\times S^1$ \cite{circle}. These states depend on the zero--mode
momenta and seems not to have a well behaved limit when the space--time
topology is taken to be trivial. 

Having found new physical states still leaves open the main difficulty
of this approach: how to extract the space--time geometric properties
from the Hilbert space. It is also necessary to compare the
non--perturbative results obtained in this approach with the
semi--classical results. We must solve these issues in the
two--dimensional models, where the problems are tractable, before
embarking in realistic four or higher dimensional gravity theories
with propagating gravitons.

\begin{acknowledgments}
This work is partially supported by FAPESP. The work of V. O. Rivelles
is partially supported by CNPq.  
\end{acknowledgments}

\begin{table}

\begin{tabular}{ccccc}
Signature & Vaccum          & Norm         & Energy   & Central Charge
\\ \hline  
positive  & $a|0>=0$          & positive definite    & positive &
$1+12\pi Q^2$ \\ 
positive  & $a^\dagger |0>=0$ & not positive definite& negative &
$-1+12\pi Q^2$ \\
negative  & $a|0>=0$          & not positive definite& positive &
$1-12\pi Q^2$ \\ 
negative  & $a^\dagger |0>=0$ & positive definite    & negative &
$-1-12\pi Q^2$ \\
\end{tabular}

\bigskip
\caption{Hilbert space norm, energy sign and central charge for all
  possible vaccum choices of a scalar field}
\label{fig1} 
\end{table}

\vskip 2cm

\begin{table}

\begin{tabular}{ccc}
Background Charges& Vaccum                    & Norm \\   \hline
$Q_a=0$           & $a_0|0>=a^\dagger_1|0>=0$ & positive definite \\
$Q_a=0$           & $a^\dagger_0|0>=a_1|0>=0$ & not positive 
definite \\
$Q_a Q^a=0$           & $a_0|0>=a^\dagger_1|0>=0$ & positive definite \\
$Q_a Q^a=0$           & $a^\dagger_0|0>=a_1|0>=0$ & not positive 
definite \\
$Q_a Q^a=-\frac{1}{6\pi}$& $a_0|0>=a_1|0>=0$      & not positive
definite \\ 
$Q_a Q^a=\frac{1}{6\pi}$& $a^\dagger_0|0>=a^\dagger_1|0>=0$& not
positive definite \\ 
\end{tabular}
\bigskip
\caption{Choices of the backgound charge and vaccum for vanishing
  central charge in the CGHS
  model}
\label{table2}
\end{table}


\begin{references}

\bibitem{CGHS} C. G. Callan, S. B. Giddings, J. A. Harvey and A. Strominger,
{\it Phys. Rev.} D{\bf 45} (1992) R1005. 

\bibitem{semiclassical} J. G. Russo, L. Susskind and L. Thorlacius,
  {\it Phys. Lett.} {\bf B292} (1992) 13; T. Banks, A. Dabholkar,
  M. Douglas and M. O'Loughlin, {\it Phys. Rev.} {\bf D45} (1992)
  3607; L. Susskind and L. Thorlacius, {\it Nucl. Phys.} {\bf B382}
  (1992) 123.

\bibitem{Cangemi-Jackiw} D. Cangemi and R. Jackiw, {\it
    Phys. Rev. Lett.} {\bf 69} (1992) 233.  

\bibitem{susy} V. O. Rivelles, {\it Phys. Lett.} {\bf B321} (1994) 189;
D. Cangemi and M. Leblanc {\it Nucl. Phys.} {\bf B420} (1994) 363;
M. M. Leite and V. O. Rivelles, {\it Class. Quantum Grav.} {\bf 12}
(1995) 627. 

\bb{Mikovic} A. Mikovi\'c, {\it Black Holes and Nonperturbative
  Canonical 2D Dilaton Gravity}, hep-th/9402095 (1994)

\bibitem{canonical} D. Cangemi and R. Jackiw, {\it Phys. Lett} {\bf
    B337} (1994) 271.

\bb{Cangemi-Jackiw-Zwiebach} D. Cangemi, R. Jackiw and B. Zwiebach,
{\it Ann. Phys.} {\bf 245} (1996) 408.

\bb{Kuchar} H. Kucha\v{r}, {\it Phys. Rev.} {\bf D39} (1989) 2263;
K. Kucha\v{r} and G. Torre, {\it J. Math. Phys.} {\bf 30} (1989) 1796. 

\bb{circle} E. Benedict, R. Jackiw and H. -J. Lee, {\it Phys. Rev}
{\bf D 54} (1996) 6213.

\bb{BRST} P. Bouwknegt, J. McCarthy and K. Pilch, {\it
  Commun. Math. Phys.} {\bf 145} (1992) 145.

\end{references}
\end{document}